\documentstyle[preprint,aps,prl,epsfig,floats,tighten]{revtex} 
\begin{document}
\draft

%
%
\title{Models of core reconstruction for the 90$^\circ$ partial
dislocation in semiconductors}
\author{R.W. Nunes$^1$ and David Vanderbilt$^2$}
\address{$^1$Departamento de F\'{\i}sica, Universidade Federal de
Minas Gerais\\ Belo Horizonte, Minas Gerais 30123-970, Brazil\\
$^2$Department of Physics and Astronomy, Rutgers University,
Piscataway, New Jersey 08854-8019\\}

\date{August 8, 2000}
\maketitle

\begin{abstract} 
We compare the models that have been proposed in the literature for
the atomic structure of the 90$^\circ$ partial dislocation in the
homopolar semiconductors, silicon, diamond, and germanium. In
particular, we examine the traditional single-period and our recently
proposed double-period core structures. {\it Ab initio} and
tight-binding results on the core energies are discussed, and the
geometries are compared in light of the available experimental
information about dislocations in these systems. The double-period
geometry is found to be the ground-state structure in all three
materials. We address boundary-conditions issues that have been
recently raised about these results. The
structures of point excitations (kinks, solitons, and kink-soliton
complexes) in the two geometries are also reviewed.
\end{abstract}


\narrowtext

\section{Introduction}

Dislocations are fundamental defects associated with the mechanisms of
plastic deformation in solids, playing also a critical role as traps
and recombination centers for carriers in semiconductors. In group IV
and III-V semicondutors, under usual regimes of plastic deformation,
60$^\circ$ and screw dislocations lying on [110] directions in \{111\}
slip planes are formed. Experimental evidence suggests these to be
dissociated into partials both at rest and in
motion~\cite{hirsch,duesbery,alexan}. The screw dissociates into two
30$^\circ$ partials and the 60$^\circ$ dislocation into a 30$^\circ$
partial and a 90$^\circ$ partial. A central issue regards bond
reconstruction at the dislocation cores, with most, if not all, of
experimental and theoretical works to date indicating that
reconstruction does take place. Knowledge of the atomic structure of
the core, and of the point-like core excitations (kinks and solitons)
which determine the modes of dislocation motion, is crucial to
understanding the mechanisms of plastic deformation at a microscopic
level.

Here, we review recent theoretical results leading to the proposal of
a new core structure for the 90$^\circ$ partial dislocation in
homopolar semiconductors~\cite{bnv,nbv3,nv}. This structure has been
called the double-period (DP) reconstruction because it involves a
doubling of the period along the dislocation direction. In the
following, we discuss this and the other models of core reconstruction
that had been previously considered for the 90$^\circ$ partial. The
traditionally accepted single-period (SP) reconstructed
core~\cite{sp90,jones,heggie,bigger,nbv1} is compared with the DP
geometry~\cite{bnv,nbv3,nv}. We also give a brief overview on the
types of point excitations occurring in the two geometries.

\section{``Quasi-fivefold'' and single-period core models}

\begin{figure}
\centerline{\epsfig{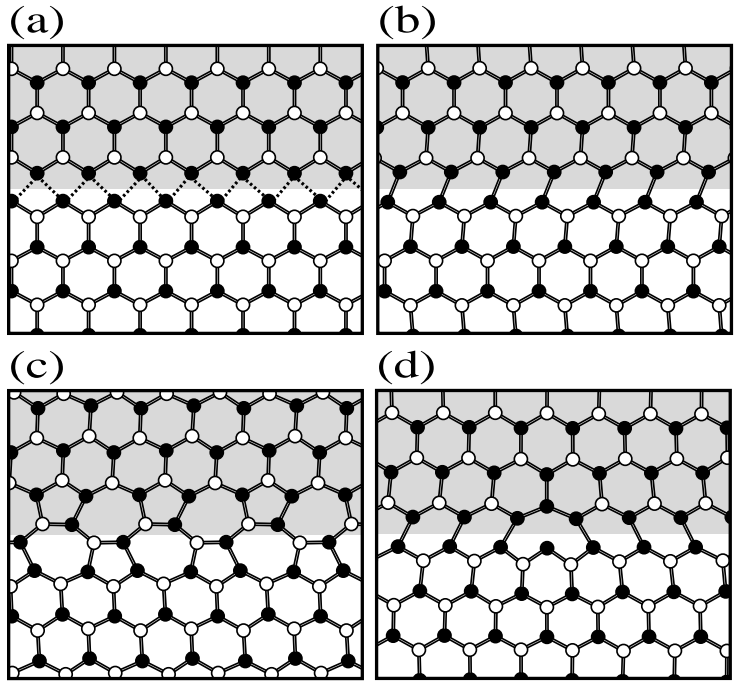}}
\vspace{0.4cm}
\caption{Models of core reconstruction of the 90$^\circ$ partial
dislocation.  (a) Symmetric QF reconstruction. (b) Symmetry-breaking
SP structure.  (c) Ground state symmetry-breaking DP structure. (d)
Soliton in the SP core.}
\label{core}
\end{figure}

Theoretical works on the atomic-structure of dislocations in
semiconductors have heavily concentrated on the 90$^\circ$ partial
dislocation~\cite{sp90,jones,heggie,bigger,nbv1}. From the
displacement field predicted for this dislocation by continuous
elasticity theory, one obtains an unreconstructed configuration, with
the dislocation core displaying a zigzag chain of
threefold-coordinated atoms running along the dislocation direction,
with broken bonds lying nearly parallel to the slip plane. Mirror
symmetry planes along the dislocation direction are present is this
configuration, as shown in figure~\ref{core}(a).  By allowing relaxation
of this core structure, while keeping the mirror symmetry planes, one
obtains a ``quasi-fivefold'' (QF) structure in which the distance between
the two zigzag chains is reduced. The quasi-fivefold denomination is
justified since this lead to atomic distances which do not
characterize full bond reconstruction~\cite{bigger}.

The way for this system to undergo full reconstruction appears
straightforward: by breaking the mirror symmetry of the
unreconstructed core, reconstructed bonds are formed in the manner
shown in figure~\ref{core}(b). Note that the lattice periodicity along
the line is preserved in this geometry, hence we shall refer to this
as the single-period structure. All dangling bonds have been
eliminated, and all the atoms are fourfold coordinated.  {\it Ab
initio} and tight-binding calculations have predicted this structure
to be substantially lower in energy (by about 0.2 eV/\AA) than the
quasi-fivefold core~\cite{bigger,nbv1}, being thus the one expected to
occur in nature.

\section{The double-period structure}

Most theoretical works on the 90$^\circ$ partial have addressed the SP
geometry. Kinks are fully reconstructed in this structure, with low
formation energies, as shown by available theoretical calculations
with values ranging from 0.1 to 0.5 eV~\cite{heggie,nbv1,nbv2,oberg}.
Kinks are thus seen to introduce little strain on the core of the SP
geometry. Since a kink and an antikink exert an attractive elastic
interaction on each other, and given the already low formation
energies above, results indicating a negligible energy for a
kink-antikink pair of minimum separation (with the kink and antikink
occurring in adjacent sites along the core) should not have been very
surprising. \"Oberg {\it et al.}  reported an energy of only 0.004 eV
for this defect, while Nunes~\cite{nunes} computed a value of 0.01
eV. Keating potential calculations on 96-atom supercells of the type
employed in our previous works~\cite{bnv,nbv1} give {\it negative} values ($\sim
-$0.3 eV) for the formation energy of the defect.

\begin{table}
\caption{Calculated energy difference in meV/\AA, between the SP- and
DP-core reconstructions of the 90$^\circ$~partial in diamond (C),
silicon, and germanium. Cell size refers to the double-period cell.
$E_{\rm DP}$ is the energy of the double-period reconstruction.  For
the single-period case, $\overline{E}_{\rm SP}$ and $\Delta E_{\rm
SP}$ are respectively the average and difference of the energies for
the two different relative arrangements of mirror symmetry-breaking.}
\smallskip
\begin{tabular}{ldddd}
 &\multicolumn{2}{c}{192-atom supercell}
 &\multicolumn{2}{c}{588-atom supercell}\\
 &$E_{\rm DP} - \overline{E}_{\rm SP}$ &$\Delta E_{\rm SP}$
 &$E_{\rm DP} - \overline{E}_{\rm SP}$ &$\Delta E_{\rm SP}$\\
\hline
C \\
\quad LDA             &$-$235   &126    \\
\quad TETB            &$-$100   & 74     &$-$76    &14     \\
\quad Keating\tablenotemark[1]
                      &$-$121   &160    \\
Si \\
\quad LDA             & $-$69    &48     \\
\quad TETB            & $-$75    &39     &$-$57    & 3     \\
\quad Keating\tablenotemark[1]
                      & $-$40    &67     \\
Ge \\
\quad LDA             & $-$58    &27     \\
\quad Keating\tablenotemark[1]
                      & $-$12    &36     \\
\end{tabular}
\tablenotetext[1]{Evaluated at LDA-relaxed structure.}
\end{table}

The full implications of these latter results went unnoticed, however,
until it was realized that continuing the process of inserting
kink-antikink pairs in the SP core would lead to a new core structure
with lower energy than the SP core itself~\cite{bnv}. In addition to
the symmetry breaking already present in the SP core, this new
structure involves a doubling of the periodicity along the dislocation
line, and for that reason it was called the double-period (DP)
reconstruction. The atomic geometry is shown in figure~\ref{core}(c).
In the original work, the DP core was introduced for silicon, being
later extended to diamond and germanium~\cite{nbv3}. In
this latter work, the DP core was found to be the ground-state in all
three materials, as shown by the {\it ab initio} and tight-binding
total energy (TBTE) results in table~I~\cite{nbv3}. Recently, Blase
{\it et al.}~\cite{blase} have shown that the DP core is more stable
in diamond over a broad range of stress states, by means of {\it ab
initio} calculations.

Examination of bond lengths and bond angles of the SP and DP structures
suggest that the DP core is able to ``pack'' the atoms more
efficiently, as indicated by the smaller average bond-length
deviations, at the expense of larger bond-angle deviations. The
balance between bond-bending and bond-stretching forces leads to the
preference of the three materials for the DP core. The difference is
rather subtle, though, as seen by the small differences in maximum
bond-length and bond-angle deviations in the two cores, as given in
table~II~\cite{nbv3}.

\section{Experimental evidence}

So far, no clear experimental evidence can decide between the two
structures. Both geometries are fully reconstructed, hence neither
gives rise to deep-gap states which would show an EPR signal. It seems
to be safely established by measurements that a rather small density
of dangling bonds is to be expected in the core of 90$^\circ$-partial
dislocation~\cite{hirsch,duesbery,alexan}. Moreover, both cores
consist entirely of fivefold, sixfold, and sevenfold rings, both being
consistent with images produced by transmission electron microscopy,
at the current level of resolution of this technique~\cite{kolar}.
Recent experimental work by Batson~\cite{batson} indicates a
DP-derived structure (called the ``Extended DP structure'' by this
author) to be more consistent with STEM and EELS experiments.

\begin{table}
\label{table2}
\caption{Minimum, maximum, and root-mean-square variations of bond
lengths and bond angles for the LDA-relaxed SP and DP structures,
relative to the corresponding bulk values.}
\smallskip
\begin{tabular}{lrrrr}
 &\multicolumn{2}{c}{bond length}
 &\multicolumn{2}{c}{bond angle}\\
 & \multicolumn{1}{c}{SP}&  \multicolumn{1}{c}{DP}
 & \multicolumn{1}{c}{SP}&  \multicolumn{1}{c}{DP} \\
\hline
C \\
\quad min          &$-$5.3\%   &$-$4.4\%   &$-$11\%    &$-$14\%     \\
\quad max          &+5.4\%   &+6.2\%   &+20\%    &+22\%     \\
\quad rms          & 3.1\%   & 2.8\%   &3.4\%    &3.6\%    \\
Si \\
\quad min          &$-$2.2\%   &$-$2.1\%   &$-$11\%    &$-$15\% \\
\quad max          &+3.0\%   &+3.5\%   &+22\%    &+23\% \\
\quad rms          & 2.6\%   & 2.3\%   &4.0\%    &4.1\% \\    
Ge \\
\quad min          &$-$2.2\%   &$-$2.1\%   &$-$11\%    &$-$15\% \\
\quad max          &+3.1\%   &+3.5\%   &+22\%    &+22\% \\
\quad rms          & 2.8\%   & 2.5\%   &4.0\%    &4.1\% \\
\end{tabular}
\end{table}

\section{The issue of boundary conditions}

Lehto and \"Oberg (LO) have recently investigated the influence of the
choice of supercell periodic boundary conditions in the relative
stability of the SP and the DP structures~\cite{lo}. Performing
supercell calculations that employed the Keating potential for Si,
they concluded that the relative stability of the SP and DP geometries
depends on the choice of boundary conditions, with the SP and DP cores
being favored for ``dipole'' and ``quadrupole'' configurations (these
are related to different possible choices of supercell boundary
conditions), respectively. Their work raises interesting
possibilities, such as the influence of neighboring extended defects,
and of the dislocation stress state (as studied in
Blase {\it et al.}~\cite{blase}) on dislocation core structure, when two
competing core structures are very nearly degenerate in energy.

In order to answer the question of which of the two structures is more
stable, in the bulk of Si at vanishing stress, we have performed
cell-size converged TBTE calculations which clearly show the DP
structure to be more stable regardless of the choice of boundary
conditions~\cite{nv}. The results are given in table~III. Note
that the numbers in this table show a bias of about $\sim$50 meV/\AA\
in the Keating potential, in favor of the SP structure.

\section{Kinks and solitons}

Symmetry breaking in the SP core gives rise to a soliton-type (S)
defect at the boundary between two stretches of the dislocation, in
which the direction of the bonds is switched, as shown in
figure~\ref{core}(d). Note the presence of a dangling bond in the core of
the defect, which explains its formation energy of 1.45 eV. Heggie and
Jones~\cite{heggie}, suggested a mechanism by which solitons in the SP
core may act as nucleating centers for kink-antikink pairs.

\begin{table}
\label{table3}
\caption{Energy of the DP relative to the SP core, in meV/\AA\ per
dislocation, for the 90$^\circ$ partial in Si.  Three different
approximations [TBTE and two different Keating parameterizations (see
Nunes and Vanderbilt~[6])] are used to compare quadrupole and
dipole boundary conditions.  Cell size refers to number of atoms in
the DP case.}
\smallskip
\begin{tabular}{lrrr}
Cell size (atoms) &192 &1440 &1920\\
\hline
TETB\\
\quad quadrupole &$-$74  &$-$56  &$-$55 \\
\quad dipole     &$-$62  &$-$52  &$-$55 \\
Keating, Ref.~[3]\\
\quad quadrupole &$-$27  & $-$1  &  1 \\
\quad dipole     &$-$14  &  3    &  5 \\
Keating, Ref.~[4]\\
\quad quadrupole &$-$22  & $-$6  & $-$5 \\
\quad dipole     &$-$13  & $-$4  & $-$3 \\
\end{tabular}
\end{table}

As discussed by Bulatov {\it al.}~\cite{bulatov}, related to the
twofold degeneracy of its reconstruction, the SP core supports two
\begin{figure}[!b]
\centerline{\epsfig{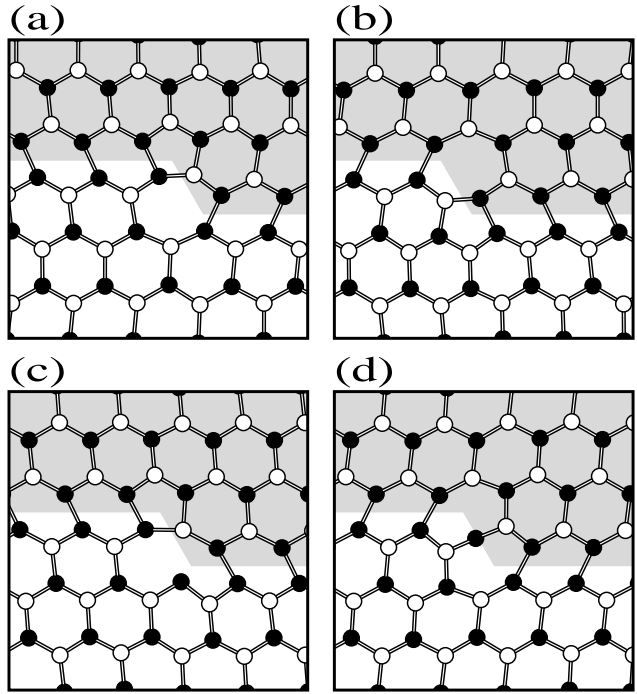}}
\vspace{0.4cm}
\caption{Core structure of kinks and soliton-kink complexes in the SP
core.  See text for notation. (a) LR kink. (b) RL kink. (c) LL complex
= LR + soliton.  (d) RR complex = RL + soliton.}
\label{SP-def}
\end{figure}
kinks and two types of kink-soliton complexes.  Referring to
figure~\ref{SP-def}(a), the reconstruction will be said to tilt to the
`left' and to the `right' on the left and right sides of the kink,
respectively.  Hence, we call this a left-right (LR) kink, the
notation following accordingly for the other kinks. In our
terminology, we call kinks the stable, fully reconstructed excitations
shown in figures~\ref{SP-def}(a) and (b), while kink-soliton complexes
are those containing a three-fold coordinated atom in their center, as
shown in figures~\ref{SP-def}(c) and~\ref{SP-def}(d). Being fully
reconstructed, the former have low formation energies (we 
obtained a value of $\sim$0.12 eV in~\cite{nbv2}), while the latter
are unstable against the emission of a soliton, in reactions of the
type $RR \rightarrow S+ LR$, as we have previoulsy
discussed~\cite{nbv1}. A detailed study of the structure and
energetics of these excitations is given in our previous
works~\cite{nbv1,nbv2}.

\begin{figure}
\centerline{\epsfig{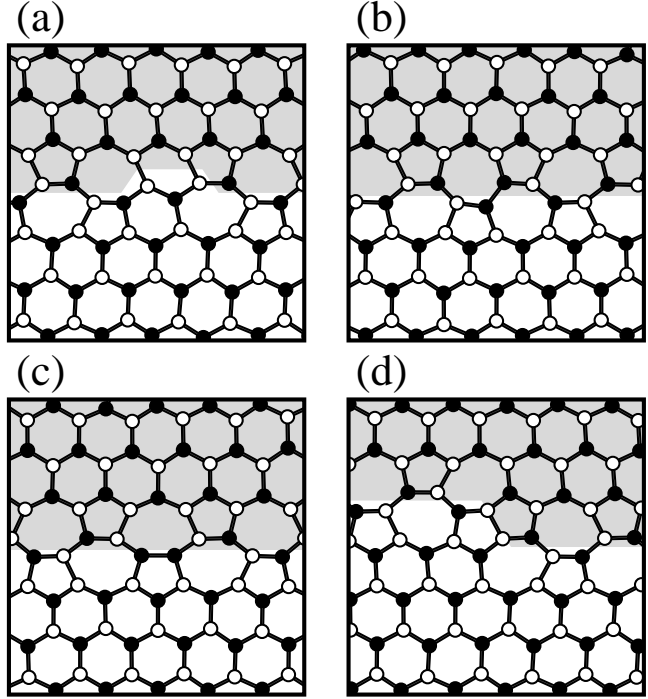}}
\vspace{0.4cm}
\caption{Examples of several types of core defects in the DP
structure.  Viewpoint is the same as for figure~1.
(a) Phase-switching defect (PSD).
(b)-(c) Direction-switching defects (DSD). (d) Kink.}
\label{DP-def}
\end{figure}

The period doubling of the DP geometry, combined with the mirror
symmetry breaking common to the two structures, gives rise to an even
richer structure of core excitations for the DP core.  Four equivalent
ground states are present(``dnqu'', ``qudn'', ``pnbu'', ``bupn'' in
the notation of Bennetto {\it et al.}~\cite{bnv}) related to each other by (110)
mirrors and by single-cell translations.  An antiphase defect occurs
at a translational domain boundary between core segments; being
referred to as a ``phase-switching defect'' (PSD)
[figure~\ref{DP-def}(a)]. As can be seen in figure~\ref{DP-def}(a), a PSD
can be regarded as a short segment of the SP structure inserted into
the DP one. Being free of dangling bonds, the PSD is expected to be a
low-energy structural excitation. A second class of defects results
from a reversal of the mirror symmetry-breaking.  These are termed
``direction-switching defects'' (DSDs); they can be classified by the
direction of switching, among other factors.  Two examples are shown
in figures~\ref{DP-def}(b) and~\ref{DP-def}(c), respectively. A DSD will necessarily
contain a dangling bond or an over-coordinated atom, so the DSDs are
expected to have higher energies than the PSDs. (The malcoordinated
atoms do not appear in figure~\ref{DP-def} as they are located just
above or below the plane of the figure.)

Finally, turning to the kink structures, because there are four
degenerate core structures to choose between on each side of the kink,
there should be at least 16 distinct kinks.  However, each of these is
paired with another into which it can be converted by displacing the
center of the kink by one lattice constant along the
dislocation~\cite{bnv,bulatov}. Altogether, 8 topologically distinct
families of kinks are found in this structure.  Furthermore, most of
these families may be classified as ``kink-defect complexes''
incorporating either a DSD, or PSD, or both, which may or may not be
energetically bound to the kink.  Those including a DSD will retain a
malcoordinated atom, and will have no reversal of the mirror
symmetry-breaking across the kink; those not including a DSD will be
fully reconstructed and will show a reversal of the mirror
symmetry-breaking.  An example of the latter kind is shown in
figure~\ref{DP-def}(d).

Much work remains to be done to characterize the point excitations of
the DP core. Some initial results on kink formation energies were
given in Bennetto {\it et al.}~\cite{bnv}, but soliton and kink
migration barriers, as well as stability of soliton-kink complexes in
this geometry, are crucial aspects determining the mechanisms of
dislocation motion, which remain to be addressed. The role of
solitonic defects in the nucleation of kink-antikink is also an
interesting possibility, to be explored in future studies.

\acknowledgments 

R.~W.~Nunes acknowledges support from CNPq-Brazil.
D.V. acknowledges support from NSF Grant DMR-99-81193.

\end{document}